\documentclass[12pt]{article}
\usepackage[utf8]{inputenc}
\usepackage[T1]{fontenc}
\usepackage{graphicx}
\usepackage{grffile}
\usepackage{longtable}
\usepackage{wrapfig}
\usepackage{rotating}
\usepackage[normalem]{ulem}
\usepackage{amsmath}
\usepackage{textcomp}
\usepackage{amssymb}
\usepackage{capt-of}
\usepackage{hyperref}
\usepackage[round]{natbib}
\usepackage{setspace}
\usepackage{authblk}
\author[1,2]{Feiming Chen}
\author[1]{Hong Laura Lu}
\author[1]{Arianna Simonetti}
\affil[1]{Center for Devices and Radiological Health, Food and Drug Administration, Silver Spring, MD, USA.}
\affil[2]{Corresponding author, E-mail: Feiming.Chen@fda.hhs.gov}
\date{}
\title{Practical Statistical Considerations for the Clinical Validation of AI/ML-enabled Medical Diagnostic Devices}
\hypersetup{
 colorlinks=true,
 urlcolor=magenta,
 citecolor=black,
 pdfauthor={Feiming Chen},
 pdftitle={Practical Statistical Considerations for the Clinical Validation of AI/ML-enabled Medical Diagnostic Devices},
 pdfkeywords={},
 pdfsubject={},
 pdfcreator={Emacs 26.1 (Org mode 9.1.9)}, 
 pdflang={English}}
\begin{document}

\maketitle

\begin{abstract}

  Artificial Intelligence (AI) and Machine-Learning (ML) models have
  been increasingly used in medical products, such as medical device
  software. General considerations on the statistical aspects for the
  evaluation of AI/ML-enabled medical diagnostic devices are discussed
  in this paper. We also provide relevant academic references and note
  good practices in addressing various statistical challenges in the
  clinical validation of AI/ML-enabled medical devices in the context
  of their intended use.

\end{abstract}


\section{Introduction}
\label{sec:org4bb7622}

Latest advances in biomedical research have made available a new
generation of medical diagnostic devices based on the promising
technology of Artificial Intelligence/Machine Learning (AI/ML) that
can be used to design advanced modelling systems. Such models are
trained on data relevant to specific tasks by optimizing certain
performance metrics, and are often embodied in software or apps
intended for medical purposes.  Sometimes a software application
functions without being part of a hardware medical
device \citep{CDRH4,SAMD}. 

  In the biomedical diagnostic field, AI/ML
may be used to develop a software algorithm for a medical intended use
with the intent to assist medical experts in making a more accurate
and rapid diagnostic decision regarding the classification, detection,
and prediction of diseases.  As part of the current trend of
automating traditional labor-intensive tasks, AI/ML technology as
applied in the biomedical field is exciting and has the potential to
provide a significant contribution to medical diagnostic systems, but
it also presents some limitations and new challenges in the
statistical aspect of clinical validation.  The term "clinical
validation" means the demonstration of objective evidence that
clinically meaningful outcomes of the medical device software can be
achieved predictively and reliably in the target population according
to the intended use of the device \citep{CDRH4,SAMD}. 

While acknowledging that the choice of clinical validation methodology
for a particular technology is generally based on factors including its
specific intended use and technological characteristics, in this
paper, we outline some general statistical considerations in the
clinical validation of AI/ML-enabled medical devices. Section
\ref{devAI} briefly describes the development of an AI model. Section
\ref{aspval} illustrates four basic aspects in the AI model
validation. Section \ref{addit} includes additional considerations for
model validation. Finally, a discussion is provided in section
\ref{discuss}.

\section{Development of an AI Model}
\label{sec:orga0d047b}
\label{devAI}

Statistical Learning Theory is a framework for AI/ML that borrows
from the field of statistics and functional analysis with the goal to
generate insight from data by using “supervised” or “unsupervised”
modeling \citep{Hastie_2009,James_2013g}, where the term
supervised/unsupervised indicates the presence/absence of a true
outcome during model development.  A simple example of a supervised
learning model is a classic linear or logistic regression.  In the following
section, we focus on supervised learning and use "AI Model" as a
shorthand umbrella term for modern predictive models or algorithms
using AI/ML technologies. An AI model is characterized by its need to
\emph{learn} from training data before it can be implemented to make
"intelligent" predictions on new, future data.

Clinical need, coupled with the availability of relevant training
data, may drive the development of an AI model \(f(X)\), that aims to
predict a target (e.g., a true clinical outcome) \(Y\) using a function
\(f\) on input \(X\).  The form of \(f\) (e.g., random forest, gradient
boosting machine, deep learning neural network) is often determined by
model developers through experience or tools like cross-validation and
may incorporate important physiological insights.  A typical AI model,
as applied to the diagnostic medical field, often includes three
parts:
\begin{enumerate}
\item A quality control algorithm that drops cases deemed as low-quality
\item A suite of models that extracts the data relationship
\item A risk classification/score algorithm with either a diagnostic or
prognostic goal, often designed to satisfy significant clinical
considerations
\end{enumerate}

Once the form of an AI model has been determined and its parameters
have been estimated from a set of training data, algorithm developers
may estimate a \emph{generalized} performance (e.g., the model
performance on out-of-sample data) using statistical techniques such
as cross-validation.  This is called \emph{internal validation}.  Since the
performance is estimated from the same training data, there is a
tendency to arrive at an over-optimistic performance estimate
\citep{simon2003pitfalls} that may not hold when the AI model is used in a
real-world situation. Two factors may account for such phenomenon:
\begin{enumerate}
\item The training data informs the model development in many ways
(e.g., cross-validation or Bootstrap adjustment
\citep{harrell1996multivariable,efron1997improvements} are often
used to tune hyperparameters or select models), and it can be
hard to get an unbiased performance estimation when the model is
already “richly” informed by the training data.
\item The \emph{generalizability} of the model performance, i.e., the
ability of the model to produce accurate predictions in a new
sample of patients, can be difficult to account for during
performance estimation.
\end{enumerate}
Given the above reasons, the internal validation provides only a
starting point on performance estimation; and the \emph{external
validation}
\citep{moons15_trans_repor_multiv_predic_model,justice99_asses_gener_progn_infor},
as explained later in section \ref{exval}, will provide a more
objective, unbiased estimate of the model performance.

\section{Basic Statistical Considerations for the Validation of a Diagnostic Device AI Model}
\label{sec:orgff439ee}
\label{aspval}

At basic level, the validation for a diagnostic device AI model
(simply referred as an AI model in the following) is no different from
how a theoretical model is validated. In math and physics, scientists
extract models from studying the real-world data (e.g., hypothesize a
law of gravity from astronomic observations) and test their predictive
powers in \emph{new} settings (e.g., use it to discover a new
planet). Similarly, in machine learning for medical applications, we
rely heavily on computers to extract models (e.g., a black box deep
learning model) from some real-world data and test their predictive
powers in \emph{new} datasets - ideally in a different geographical and
clinical setting. In this scenario, a key validation concern is the
\emph{strength of the external evidence} that is used to support the
\emph{generalizability} (a.k.a. "external validity", an ability to provide
accurate predictions in a new sample of patients
\citep{justice99_asses_gener_progn_infor}) of an AI model performance.
This is discussed in section \ref{exval} below.

\subsection{Four Aspects for the Validation of an AI Model}
\label{sec:org611bafa}
Four basic aspects for the validation of an AI model are 
as follows:
\begin{enumerate}
\item Quality of the Training Data
\item Quality of the Validation Data
\item Model Precision (i.e., repeatability and reproducibility)
\item Model Accuracy
\end{enumerate}
For general reference, the readers can find some basic recommendations
for AI-enabled medical products in
\citep{liu2020reporting,rivera2020guidelines}, a guideline for reporting
diagnostic accuracy studies (STARD) \citep{cohen2016stard}, a checklist
for predictive model validation (TRIPOD)
\citep{moons15_trans_repor_multiv_predic_model}, and a helpful list of
fundamental statistical considerations in the context of biomarker
test in \citep{de2013evaluation}.  In section \ref{addit}, additional
statistical considerations are provided.

\subsection{Quality of the Training Data}
\label{sec:org7b76899}

One reason to assess the adequacy of the training data is to guard
against the \emph{AI bias}, a potential tendency to miss prediction target
in systematic, but often unforeseeable way, which has been
well-documented \citep{hao2019ai}. Although the AI bias may be hard to
eliminate completely, a good practice is to ensure that the training
data sufficiently represents the intended use (target) population of a
medical diagnostic device, often in terms of relevant disease
conditions (e.g., the full spectrum of disease severity) and
demographics (e.g., age, sex, race, ethnicity, geography), to avoid
building bias into a model. A descriptive statistical analysis on the
various distributions of the training data, as well as a subgroup
analysis \citep{rosenkranz2020exploratory} on the homogeneity of the
model performance, can help check the alignment of the training data
with the target population and possibly prevent or reduce AI bias.

Some common sense checks may also help improve the quality of the
training data.  For example, in image recognition tasks, fatal
confounding may occur when all the diseased cases are imaged with the
same instrument that is identified by a label on the image.  Such
confounding in the training data, if not identified and mitigated, may
be opportunistically exploited by an AI model. Thus, it is important
that meaningful, unbiased, representative training data is used so
that the trained model is less likely to be overfitted to the
potential biases and idiosyncrasies of the training data and is more
likely to generalize and maintain performance with a new real-world
data.

\subsection{Quality of the Validation Data}
\label{sec:orga16c474}
\label{exval}

A common pitfall in the selection of validation/test data for
validating an AI model is that the test data resembles the training
data "too well" \citep{altman2000we}.  For example, this can happen when
a random splitting of a procured database is used to generate the
training and test data. Another example is when both the training and
test data are coming from the same clinical site or health care
system. In both cases, the performance of the AI model could be
overestimated since the model was trained on data similar to the test
data.  In other words, the test data in both cases is lacking the
\emph{uncontrollable variability} that is expected in a real-world setting
but \emph{unknown} to the trained model.

A significant challenge is that both the training and test data are
expected to represent the intended use population of a device, with
minimal overlap (in terms of location and time) between test data and
training data.  This is because the real validation goal is to check
the \emph{generalizability}, also called \emph{external validity}, of the model
performance: whether a model trained using a particular data source
can maintain its claimed predictive performance in a different
real-world situation that the model has not seen before. A good
reference on generalizability is
\citep{justice99_asses_gener_progn_infor}.

Given the above consideration and the validation goal, a good practice
is to perform an \emph{external validation}
\citep{justice99_asses_gener_progn_infor} where the test data are
\emph{independent validation data}.  Such test data could come from diverse
health care systems with diverse clinical practices in diverse
geographies (multi-site), an expectation that could also apply to the
training data used in the development phase of the model.  Although
ideally the training data would come from a similarly diverse but
\emph{different} source, an "overlap" may often exist due to feasibility or
cost constraints. In such case, alternative forms of
"transportability" \citep{justice99_asses_gener_progn_infor} for
validation may be considered.

\subsection{Model Precision: Repeatability and Reproducibility}
\label{sec:org1f218f5}

A precision study gauges the variability of a device output when
making repeated measurements on the same patient, either with the
same operator and product (\emph{repeatability}), or with different operator
and product (\emph{reproducibility}).
More generally, repeatability is the closeness of agreement of
repeated measurements taken under the same conditions; and
reproducibility is the closeness of agreement of repeated measurements
taken under different, pre-specified conditions \citep{CLSI}. 

Key statistics to summarize the repeatability/reproducibility, based
on a variance component analysis using a model’s continuous metric
(e.g., a probability score), are the standard deviation (SD) and the
percent coefficient of variation (\%CV). If the SD or \%CV is too large,
then additional engineering improvement to the AI model may be appropriate.
Improving the AI model to reduce SD or \%CV provides a low-cost way to
improve product quality and the success likelihood of a future pivotal
clinical study. This is, in part, because the clinical reference
standard (i.e., the best available method for establishing the
presence of absence of the target condition \citep{cohen2016stard}) does not need to be
measured in a precision study.
Depending on the product, additional factors may be considered in the
precision study. In image classification tasks, an AI model may be
sensitive to data perturbation (e.g., image translation/rotation,
light intensity change, random noise). Such sensitivity could be
abundant for an AI-enabled medical device software running on a
generic smartphone using its camera to capture measurement data (e.g.,
skin lesion analyzers). Recently, an interesting type of data
perturbation that may flip the AI model result is documented in
\citep{finlayson2019adversarial}, where the basic idea is that
imperceptible data/image manipulation may be designed to completely
fail an AI model.

To ensure an AI model prediction is robust, algorithm developers often
use \emph{data augmentation} techniques to help increase the model
robustness by adding multiple variations of the same data to the
training set during the model development stage, in the expectation
that a model so trained would withstand "stress testing" during the
model validation stage. In a sense, a basic type of "stress test" is
the precision study mentioned above, which can check a model’s
robustness with repetition of the operating procedure on the same
patient when feasible.

\subsection{Model Accuracy}
\label{sec:orgcb22a9b}

The key performance assessment of an AI model is its diagnostic
accuracy, which is evaluated in a pivotal performance study. Due to
sampling variation, the uncertainties of the accuracy estimates are
typically quantified, usually in the form of 95\% confidence
intervals. The study acceptance criteria can be based on statistical
inferences using hypothesis testing methods (e.g., comparing a
lower/upper confidence limit to a pre-specified performance goal).  It
is often very useful to have a comparator that can be tested and
evaluated on the same patient/data as the device.  This comparator can
be the clinician, another device, or standard of care. The evaluation
on the same patient/data is key to mitigate differences in the task difficulty
levels and disease spectrum due to sampling variation. 

Depending on the nature of the diagnostic output (i.e., binary,
polychotomous, or continuous), different evaluation metrics are
possible.
\begin{itemize}
\item For binary diagnostic output, sensitivity, specificity,
positive/negative predictive values (PPV/NPV), and positive/negative
diagnostic likelihood ratios (LR+/LR-) may be the preferred
performance measures to be evaluated \citep{CDRH2}.
\item For risk stratification output that classifies a patient into one of
multiple risk groups and that may often be found in prognostic
models \citep{altman2000we,altman2009prognosis}, pre/post-test risks
and diagnostic likelihood ratios may be used.
\item For risk score output that evaluates a patient’s disease risk with a
continuous probability, calibration plot, ROC curve, and decision
curve analysis may be used \citep{steyerberg2014towards} (also see
section \ref{risk-score-validation} below).  In the context of
biomarker evaluation, the predictiveness curve analysis
\citep{pepe2008integrating,huang2009parametric,huang2007evaluating} may
be used.
\item For continuous score, agreement study methods using scatter plot,
Deming regression \citep{deming1943statistical}, and Bland-Altman
analysis \citep{bland1999measuring} are often used.
\end{itemize}

\section{Other Statistical Considerations for the Validation of a Diagnostic Device AI Model}
\label{sec:org77f4a07}
\label{addit}
\subsection{Study Design Considerations}
\label{sec:org59a1ecd}
The objective evaluation of a diagnostic device typically involves an
external validation study (a.k.a. diagnostic performance study) to be
performed with the locked-down, fixed, market-ready version of the
device, using subjects completely different and separate from those
used in the training of the AI model and in all prior clinical
investigations \citep{de2013evaluation,CDRH1}.  In the context of AI
models, guiding principles have also been developed to promote Good
Machine Learning Practice (GMLP) \citep{GMLP}.

A typical diagnostic performance study has a non-randomized
single-arm comparative study design that compares the subject device
with either a clinical reference standard or a comparator device.
Note some SaMD's are aids to the clinicians (e.g., in radiology
applications).  In such case, instead of a standalone performance
study, a reader study (where the reader performs a task with and
without a SaMD) is often used to compare the performance with the SaMD
and the performance without the SaMD.

An appropriate study design and study conduct
\citep{CDRH1,pepe2008pivotal} for an external validation study reduces
\emph{bias} and builds confidence in the study results.  Here are some
examples of good study design and study conduct practices:
\begin{itemize}
\item Study Design \citep{CDRH1}:
\begin{itemize}
\item The test data (independent validation data) is representative of
the intended use population of the device.
\item Pre-specified clinical study protocol and statistical analysis
plan: pre-specification is used to avoid post-hoc analysis that
may bias the performance results.
\item A \emph{prospective} design, or if not feasible, a
\emph{prospective-retrospective} design \citep{de2013evaluation}. Clinical
data is best collected prospectively in the intended use
population according to the intended use of the device.  A
prospective-retrospective design allows an existing data source to
be sampled as if in a prospective manner (e.g., sampling design is
determined before unmasking the data source).
\item An adequate \emph{masking} protocol, so that knowledge of one result
(e.g., clinical reference standard) does not influence the other
paired result (e.g., device result).
\item Demonstration of control of type I and type II errors: the former ensures
an observation of study success is not due to an outsized (e.g.,
more than 5\%) false positive error during statistical hypothesis
testing; the latter ensures reasonable study success likelihood
(e.g., more than 80\%) with adequate sample size, assuming
effectiveness of the device.
\end{itemize}
\item Study Conduct:
\begin{itemize}
\item Algorithm developers are masked from the test data. This ensures
no information from the independent validation data would bias the
algorithm development.  In the external validation study,
interpreters of the device/test outputs are masked from the reference results
(ground truth) and vice versa.
\item "Freeze/lock" the algorithm before test data collection or
unmasking. This ensures the algorithm will not be tweaked
opportunistically in light of the independent validation data to
bias the performance result.
\end{itemize}
\end{itemize}

\subsection{Influence of Signal Quality Control Algorithm}
\label{sec:org12f5d85}
An AI model is often preceded by a quality control algorithm that
discards "low" quality cases (e.g., images that are blurred or
botched) from further processing. However, such low-quality cases may
sometimes \emph{confound} with truly hard/difficult ones --- an example of
\emph{missing not at random} (MNAR), which may lead to diagnostic
performance that is improved in accuracy metrics (e.g., sensitivity
and specificity) but may also be more biased (e.g., in the sense that
more patients than warranted may not get any results due to declared
low-quality events).

For example, compare two hypothetical AI-enabled diagnostic
devices (A and B) using cellphone cameras for certain skin disease
detection. Assume they use the same diagnostic algorithms, except that
A has a more aggressive quality control (QC) algorithm than B in
declaring low-quality cases. After excluding those cases that fail the
QC algorithm, it may not be surprising to observe that A would have a
better diagnostic performance than B, because many low-quality images
dropped by A but not by B may actually be good quality but difficult
cases which are not included in the performance assessment for A.

Thus, a good practice is to examine the influence of a QC algorithm by
checking the proportion of low-quality dropouts and the result of a
sensitivity analysis assuming a worst-case scenario (i.e., assuming
the QC failure cases are all difficult ones that the AI model fails to
classify successfully).  Table~\ref{tab:org340c9e9} is a mock-up confusion table that
illustrates a potential evaluation of the impact of QC failures
(labeled as “Ungradable”) on diagnostic performance.

\begin{table}[htbp]
\caption{\label{tab:org340c9e9}
Evaluation of QC Failures in a Confusion Table}
\centering
\footnotesize
\begin{tabular}{lll|ll}
Device & True & True & Post-test Risk & Likelihood\\
Result & Positive & Negative &  & Ratio\\
\hline
Positive & A & D & \(\frac{A}{A+D}\) & \(\frac{A/(A+B+C)}{D/(D+E+F)}\)\\
 &  &  &  & \\
Negative & B & E & \(\frac{B}{B+E}\) & \(\frac{B/(A+B+C)}{E/(D+E+F)}\)\\
 &  &  &  & \\
Ungradable & C & F & \(\frac{C}{C+F}\) & \(\frac{C/(A+B+C)}{F/(D+E+F)}\)\\
\hline
Worst Case & Sensitivity & Specificity & Pre-test Risk & 1\\
Scenario & \(=\frac{A}{A+B+C}\) & \(=\frac{E}{D+E+F}\) & \(=\frac{A+B+C}{A+B+C+D+E+F}\) & \\
\end{tabular}
\end{table}

\subsection{Validation of AI Model with Risk Score Output}
\label{sec:org34a42bb}
\label{risk-score-validation}

Certain clinical considerations (e.g., long-term trend monitoring)
may justify an AI model producing a continuous risk score that
reflects the likelihood of a physiological event (e.g., a cardiac or
hemodynamic event in a time window) and that may be continuously
updated over time. The validation of such risk score may call for a
larger sample size than usual to reduce uncertainty in the evaluation.

In general, two aspects of the accuracy evaluation are \emph{calibration}
(i.e., how well does a predicted probability match the observed
frequency in the study population) and \emph{discrimination} (i.e., how
well do the rankings of predicted scores match those of observed
frequencies in the study population)
\citep{justice99_asses_gener_progn_infor}. In addition, some
evaluations for clinical usefulness such as the decision curve
analysis
\citep{vickers2006decision,vickers2019simple,fitzgerald2015decision},
relative utility curves \citep{baker2009using}, standardized net benefits
\citep{pepe2016early,marsh2020statistical} may be helpful.

In the model development stage, it is a good practice to check and
improve the model calibration \citep{chen2018calibration}. A calibration
evaluation \citep{steyerberg2014towards,van2016calibration} may involve
the estimation of calibration-in-the-large (i.e., the intercept that
compares the mean of all predicted risks with the mean observed risk),
calibration-slope (reflecting the extent of shrinkage), and
calibration plot.  A basic calibration analysis can be done with a
logistic regression of disease status on the logit of the risk
prediction. It is called the logistic recalibration method or Cox
calibration analysis \citep{miller1993validation,calster2014calibration,cox1958two}.

An AI model may be built using training data with \emph{high} disease
prevalence (e.g., 50/50 split between positive and negative
cases). Thus, the prevalence of the training data may be quite
different from that of the validation data (which would mirror the
target population and may have very low disease prevalence). This
would cause a problem of calibration-in-the-large and could be fixed
with \emph{prevalence scaling} \citep{horsch2008prevalence}, which uses Bayes
theorem to adjust the prediction probability in light of the realistic
prevalence expectation in a target population.  Such issue illustrates
the challenge of validating risk scores, a validation that may call
for a lot more data to train and test.

A discrimination evaluation may involve the estimation of ROC curve,
as well as a \emph{c} index \citep{harrell1984regression}, which reflects the
concordance between prediction probabilities and actual outcomes and
is equivalent to the area under the curve for a ROC curve (ROC-AUC).
The discriminatory power of the model can be evaluated by comparing
the \(c\) index/ROC-AUC to a pre-specified performance goal or the
performance of a comparator (e.g., standard of care).

Since a risk score may be subject to thresholding by clinicians at
different levels, it may be helpful to examine the diagnostic
performance (e.g., sensitivity/specificity and their 95\% confidence
intervals) at potential clinically relevant thresholds (e.g., every
decile or a finer grid). 

A clinical usefulness evaluation may involve a decision curve analysis
or other tools. A decision curve \citep{vickers2006decision} visualizes a
type of risk-adjusted benefit (a.k.a. net benefit) at varying levels
of risk tolerance, where the risk tolerance (a.k.a. threshold
probability) is between zero (no tolerance for false negative) and one
(no tolerance for false positive), and the risk-adjusted benefit is
quantified by the number of true positives subtracted by the relative
cost of false positives. This type of analysis could facilitate
comparison among different diagnostic alternatives and inform
population-level health policies.  Note that sometimes an appropriate
net benefit analysis depends on whether a rule-in test (e.g., when a
disease needs to be confirmed and the treatment is harmful) or a
rule-out test (e.g., when a disease is not to be missed and the
treatment is safe) may be clinically needed.

To further understand how a risk score would influence the clinical
decision, it may be possible to compute a kind of diagnostic
likelihood ratio based on the continuous risk score, which can then be
used to update a patient’s pre-test disease probability to get a
post-test disease probability using Bayes’ theorem
\citep{sox2013medical}. More research in this area
\citep{ost2022interpretation} may be helpful.

Since the validation of a risk score may call for a large sample size
and involves extensive work in evaluating calibration, discrimination,
and clinical decision making, a practical alternative is to predict
the risk with a few risk strata (i.e., risk stratification) instead of
a continuous risk score.  A risk stratification output may often be
derived by thresholding a risk score at different levels during the
model development. Risk thresholds that determine the clinical action
(or inaction) are sometimes available from clinical guidelines
\citep{pepe2016early}.  Upon a satisfactory statistical evaluation using
post-test risks and diagnostic likelihood ratios across all strata
under an external validation study, risk stratification can often
provide better clarity to the clinical utility of an AI model than a
risk score.

\subsection{Time-to-Event Outcomes in AI-enabled Diagnostic Device Validation Data}
\label{sec:org2a624f5}
\label{time2event}

An AI model may have a prognostic output that predicts whether a
future event may happen (e.g., likelihood of cardiac mortality in 10
years).  Such a model (e.g., Cox regression) may be derived by adding
new innovative biomarkers to baseline covariates (e.g., age, sex) and
common clinical risk factors (e.g., smoking status, diabetes,
hypertension).  Validating the AI model may involve collecting
time-to-event outcomes along with missing data status (e.g., due to
patient withdrawal or loss to follow-up).  Calibration and
discrimination analysis may be done via risk stratification
\citep{royston2013external}.  In addition, the contribution of new
biomarkers, in terms of the fraction of prognostic information that is
new information, may be quantified \citep{harrell-added-value}.  An
example analysis is described in \citep{royston2013external} and may be
applicable beyond the Cox model.  In general, we may use the survival
analysis techniques in the following way:
\begin{itemize}
\item Form risk groups from the prognostic output and compare their
Kaplan-Meier curves. Post-test risks across risk groups at any
specific time (e.g., one year after start) can be compared using
corresponding Kaplan-Meier estimates and their 95\% confidence
intervals.
\item Check calibration by comparing Kaplan-Meier estimates to apparent
risk levels. For example, a risk group with an average apparent risk
level of 80\% based on the prognostic outputs therein is expected to
roughly match its corresponding Kaplan-Meier estimate.
\item Check the contribution of new biomarkers (e.g., in a Cox model or
Andersen-Gill model, the latter of which is an extension of the Cox
proportional hazard model suitable for analysis of recurrent events)
with likelihood ratio test (LRT) and visualizations (e.g., compare
histograms of predicted probabilities between the baseline and the
full model) \citep{harrell-added-value}.
\end{itemize}

\subsection{Alternative to AI Black Box Models}
\label{sec:org95fdc4a}

An AI model may display a degree of opacity or may be a black box that
defies interpretation.  An opaque or black box model typically involves a
\emph{higher} level of validation evidence than \emph{interpretable} models, in
terms of the representativeness and the size of the
training/validation data, partly because a modern AI model such as a
deep learning neural network can \emph{shatter} (i.e., memorize) any
(small) training data, even if they are completely random
labels/noises \citep{zhang2021understanding}.

It is desirable that a simple and interpretable model could be used to
approximate or replace a black box model, and there is research
suggesting that an interpretable model built with advanced methods may
have similar predictive performance as a black box model
\citep{rudin2019stop}.

\subsection{Improving Internal Validation Quality of AI Models}
\label{sec:orgaf15242}

Algorithm developers often update estimated performance throughout
iterative improvement of an AI model before they decide to go into a
new clinical trial. A risk during the model development stage is that
a model may be over-trained, leading to an unrealistic performance
expectation. It is possible to avoid potential performance
over-optimism with a high-quality internal validation test data that
would resemble an external validation test data and inform a more
realistic performance expectation. In addition, it may be possible to
avoid over-training a model, due to repeated use of the test data, by
considering a \emph{privacy-preserving} data analysis
\citep{dwork2015reusable,gossmann2021test}, which perturbs the test data
with small random noise whenever it is accessed for performance
assessment. This would prevent a potentially over-trained model from
performing well on each new version of the randomly perturbed test
data. More research is needed here to inform better practice in
diagnostic device area.

\section{Future Challenges}
\label{sec:org260e472}
\label{discuss}

The field of AI-enabled diagnostic applications is expanding rapidly,
which also brings more statistical challenges regarding their clinical
validations \citep{friedrich2021there}.  As an example, in the future,
the intended use of an AI-enabled diagnostic device may include a
capability for continuous learning (i.e., ongoing model training) from
real-world data, which would presumably improve its diagnostic
accuracy over time to benefit users \citep{lee2020clinical}. However, the
validation of such device may be very challenging.  Another
challenging example is that the clinical reference standard may not be
available during the external validation study for an innovative AI
model (e.g., a prognostic biomarker for impending critical care need,
intended for continuous monitoring of patients in critical
conditions).

Despite all the potential challenges for validating an AI/ML-enabled
medical diagnostic device, we hope readers of this article can get a
better understanding of various validation aspects, follow sound
statistical principles, apply reasonable statistical methods, and gain
confidence as practitioners over time.

\section{Acknowledgments}
\label{sec:orga0a5763}
The authors wish to acknowledge the following people whose comments
led to a substantially improved paper: Rajesh Nair, Felipe Aguel,
Lilly Yue, Yunling Xu, Nicholas Petrick, Gene Pennello, Matthew
Diamond, Kathryn Drzewiecki.

\section{Disclosure statement}
\label{sec:orgda21d2a}
The authors report there are no competing interests to declare.
\section{Funding}
\label{sec:orgf39becd}
The authors report there is no funding associated with the work featured in this article.

\bibliographystyle{plainnat}
\bibliography{ref}

\begin{thebibliography}{52}
\providecommand{\natexlab}[1]{#1}
\providecommand{\url}[1]{\texttt{#1}}
\expandafter\ifx\csname urlstyle\endcsname\relax
  \providecommand{\doi}[1]{doi: #1}\else
  \providecommand{\doi}{doi: \begingroup \urlstyle{rm}\Url}\fi

\bibitem[Altman and Royston(2000)]{altman2000we}
Douglas~G Altman and Patrick Royston.
\newblock What do we mean by validating a prognostic model?
\newblock \emph{Statistics in medicine}, 19\penalty0 (4):\penalty0 453--473,
  2000.

\bibitem[Altman et~al.(2009)Altman, Vergouwe, Royston, and
  Moons]{altman2009prognosis}
Douglas~G Altman, Yvonne Vergouwe, Patrick Royston, and Karel~GM Moons.
\newblock Prognosis and prognostic research: validating a prognostic model.
\newblock \emph{Bmj}, 338, 2009.

\bibitem[Baker et~al.(2009)Baker, Cook, Vickers, and Kramer]{baker2009using}
Stuart~G Baker, Nancy~R Cook, Andrew Vickers, and Barnett~S Kramer.
\newblock Using relative utility curves to evaluate risk prediction.
\newblock \emph{Journal of the Royal Statistical Society: Series A (Statistics
  in Society)}, 172\penalty0 (4):\penalty0 729--748, 2009.

\bibitem[Bland and Altman(1999)]{bland1999measuring}
J~Martin Bland and Douglas~G Altman.
\newblock Measuring agreement in method comparison studies.
\newblock \emph{Statistical methods in medical research}, 8\penalty0
  (2):\penalty0 135--160, 1999.

\bibitem[Calster and Steyerberg(2014)]{calster2014calibration}
Ben~Van Calster and Ewout~W Steyerberg.
\newblock Calibration of prognostic risk scores.
\newblock \emph{Wiley StatsRef: Statistics Reference Online}, pages 1--10,
  2014.

\bibitem[Cen(2007)]{CDRH2}
\emph{Statistical Guidance on Reporting Results from Studies Evaluating
  Diagnostic Tests}.
\newblock Center for Devices and Radiological Health, US Food and Drug
  Administration, final edition, March 2007.

\bibitem[Cen(2013)]{CDRH1}
\emph{Design Considerations for Pivotal Clinical Investigations for Medical
  Devices}.
\newblock Center for Devices and Radiological Health, US Food and Drug
  Administration, final edition, November 2013.

\bibitem[Cen(2017)]{CDRH4}
\emph{Software as a Medical Device (SAMD): Clinical Evaluation}.
\newblock Center for Devices and Radiological Health, US Food and Drug
  Administration, December 2017.

\bibitem[Chen et~al.(2018)Chen, Sahiner, Samuelson, Pezeshk, and
  Petrick]{chen2018calibration}
Weijie Chen, Berkman Sahiner, Frank Samuelson, Aria Pezeshk, and Nicholas
  Petrick.
\newblock Calibration of medical diagnostic classifier scores to the
  probability of disease.
\newblock \emph{Statistical methods in medical research}, 27\penalty0
  (5):\penalty0 1394--1409, 2018.

\bibitem[Cli(2022)]{CLSI}
\emph{Harmonized Terminology Database}.
\newblock Clinical and Laboratory Standards Institute (CLSI), 2022.
\newblock https://htd.clsi.org.

\bibitem[Cohen et~al.(2016)Cohen, Korevaar, Altman, Bruns, Gatsonis, Hooft,
  Irwig, Levine, Reitsma, De~Vet, et~al.]{cohen2016stard}
J{\'e}r{\'e}mie~F Cohen, Dani{\"e}l~A Korevaar, Douglas~G Altman, David~E
  Bruns, Constantine~A Gatsonis, Lotty Hooft, Les Irwig, Deborah Levine,
  Johannes~B Reitsma, Henrica~CW De~Vet, et~al.
\newblock Stard 2015 guidelines for reporting diagnostic accuracy studies:
  explanation and elaboration.
\newblock \emph{BMJ open}, 6\penalty0 (11):\penalty0 e012799, 2016.

\bibitem[Cox(1958)]{cox1958two}
David~R Cox.
\newblock Two further applications of a model for binary regression.
\newblock \emph{Biometrika}, 45\penalty0 (3/4):\penalty0 562--565, 1958.

\bibitem[De et~al.(2013)De, Meier, Tang, Li, Gwise, Gomatam, and
  Pennello]{de2013evaluation}
Arkendra De, Kristen Meier, Rong Tang, Meijuan Li, Thomas Gwise, Shanti
  Gomatam, and Gene Pennello.
\newblock Evaluation of heart failure biomarker tests: a survey of statistical
  considerations.
\newblock \emph{Journal of cardiovascular translational research}, 6\penalty0
  (4):\penalty0 449--457, 2013.

\bibitem[Deming(1943)]{deming1943statistical}
William~Edwards Deming.
\newblock \emph{Statistical Adjustment of Data}.
\newblock wiley, 1943.

\bibitem[Dwork et~al.(2015)Dwork, Feldman, Hardt, Pitassi, Reingold, and
  Roth]{dwork2015reusable}
Cynthia Dwork, Vitaly Feldman, Moritz Hardt, Toniann Pitassi, Omer Reingold,
  and Aaron Roth.
\newblock The reusable holdout: Preserving validity in adaptive data analysis.
\newblock \emph{Science}, 349\penalty0 (6248):\penalty0 636--638, 2015.

\bibitem[Efron and Tibshirani(1997)]{efron1997improvements}
Bradley Efron and Robert Tibshirani.
\newblock Improvements on cross-validation: the 632+ bootstrap method.
\newblock \emph{Journal of the American Statistical Association}, 92\penalty0
  (438):\penalty0 548--560, 1997.

\bibitem[FDA(2021)]{GMLP}
\emph{Good Machine Learning Practice for Medical Device Development:Guiding
  Principles}.
\newblock FDA, Health Canada, MHRA, October 2021.

\bibitem[Finlayson et~al.(2019)Finlayson, Bowers, Ito, Zittrain, Beam, and
  Kohane]{finlayson2019adversarial}
Samuel~G Finlayson, John~D Bowers, Joichi Ito, Jonathan~L Zittrain, Andrew~L
  Beam, and Isaac~S Kohane.
\newblock Adversarial attacks on medical machine learning.
\newblock \emph{Science}, 363\penalty0 (6433):\penalty0 1287--1289, 2019.

\bibitem[Fitzgerald et~al.(2015)Fitzgerald, Saville, and
  Lewis]{fitzgerald2015decision}
Mark Fitzgerald, Benjamin~R Saville, and Roger~J Lewis.
\newblock Decision curve analysis.
\newblock \emph{Jama}, 313\penalty0 (4):\penalty0 409--410, 2015.

\bibitem[Friedrich et~al.(2021)Friedrich, Antes, Behr, Binder, Brannath,
  Dumpert, Ickstadt, Kestler, Lederer, Leitg{\"o}b, et~al.]{friedrich2021there}
Sarah Friedrich, Gerd Antes, Sigrid Behr, Harald Binder, Werner Brannath,
  Florian Dumpert, Katja Ickstadt, Hans~A Kestler, Johannes Lederer, Heinz
  Leitg{\"o}b, et~al.
\newblock Is there a role for statistics in artificial intelligence?
\newblock \emph{Advances in Data Analysis and Classification}, pages 1--24,
  2021.

\bibitem[Gossmann et~al.(2021)Gossmann, Pezeshk, Wang, and
  Sahiner]{gossmann2021test}
Alexej Gossmann, Aria Pezeshk, Yu-Ping Wang, and Berkman Sahiner.
\newblock Test data reuse for the evaluation of continuously evolving
  classification algorithms using the area under the receiver operating
  characteristic curve.
\newblock \emph{SIAM Journal on Mathematics of Data Science}, 3\penalty0
  (2):\penalty0 692--714, 2021.

\bibitem[Hao(2019)]{hao2019ai}
Karen Hao.
\newblock This is how ai bias really happens--and why it’s so hard to fix.
\newblock \emph{MIT Technology Review}, 2019.

\bibitem[Harrell(2013)]{harrell-added-value}
Frank Harrell.
\newblock \emph{Statistically Efficient Ways to Quantify Added Predictive Value
  of New Measurements}, 2013.
\newblock https://www.fharrell.com/post/addvalue.

\bibitem[Harrell~Jr et~al.(1984)Harrell~Jr, Lee, Califf, Pryor, and
  Rosati]{harrell1984regression}
Frank~E Harrell~Jr, Kerry~L Lee, Robert~M Califf, David~B Pryor, and Robert~A
  Rosati.
\newblock Regression modelling strategies for improved prognostic prediction.
\newblock \emph{Statistics in medicine}, 3\penalty0 (2):\penalty0 143--152,
  1984.

\bibitem[Harrell~Jr et~al.(1996)Harrell~Jr, Lee, and
  Mark]{harrell1996multivariable}
Frank~E Harrell~Jr, Kerry~L Lee, and Daniel~B Mark.
\newblock Multivariable prognostic models: issues in developing models,
  evaluating assumptions and adequacy, and measuring and reducing errors.
\newblock \emph{Statistics in medicine}, 15\penalty0 (4):\penalty0 361--387,
  1996.

\bibitem[Hastie et~al.(2009)Hastie, Tibshirani, and Friedman]{Hastie_2009}
Trevor Hastie, Robert Tibshirani, and Jerome Friedman.
\newblock The elements of statistical learning.
\newblock \emph{Springer Series in Statistics}, 2009.
\newblock ISSN 2197-568X.
\newblock \doi{10.1007/978-0-387-84858-7}.
\newblock URL \url{http://dx.doi.org/10.1007/978-0-387-84858-7}.

\bibitem[Horsch et~al.(2008)Horsch, Giger, and Metz]{horsch2008prevalence}
Karla Horsch, Maryellen~L Giger, and Charles~E Metz.
\newblock Prevalence scaling: Applications to an intelligent workstation for
  the diagnosis of breast cancer.
\newblock \emph{Academic radiology}, 15\penalty0 (11):\penalty0 1446--1457,
  2008.

\bibitem[Huang and Pepe(2009)]{huang2009parametric}
Ying Huang and MS~Pepe.
\newblock A parametric roc model-based approach for evaluating the
  predictiveness of continuous markers in case--control studies.
\newblock \emph{Biometrics}, 65\penalty0 (4):\penalty0 1133--1144, 2009.

\bibitem[Huang et~al.(2007)Huang, Sullivan~Pepe, and Feng]{huang2007evaluating}
Ying Huang, Margaret Sullivan~Pepe, and Ziding Feng.
\newblock Evaluating the predictiveness of a continuous marker.
\newblock \emph{Biometrics}, 63\penalty0 (4):\penalty0 1181--1188, 2007.

\bibitem[Int(2017)]{SAMD}
\emph{Software as a Medical Device (SaMD): Clinical Evaluation}.
\newblock International Medical Device Regulators Forum, September 2017.

\bibitem[James et~al.(2013)James, Witten, Hastie, and Tibshirani]{James_2013g}
Gareth James, Daniela Witten, Trevor Hastie, and Robert Tibshirani.
\newblock An introduction to statistical learning.
\newblock \emph{Springer Texts in Statistics}, 2013.
\newblock ISSN 2197-4136.
\newblock \doi{10.1007/978-1-4614-7138-7}.
\newblock URL \url{http://dx.doi.org/10.1007/978-1-4614-7138-7}.

\bibitem[Justice(1999)]{justice99_asses_gener_progn_infor}
Amy~C. Justice.
\newblock Assessing the generalizability of prognostic information.
\newblock \emph{Annals of Internal Medicine}, 130\penalty0 (6):\penalty0 515,
  1999.
\newblock \doi{10.7326/0003-4819-130-6-199903160-00016}.
\newblock URL \url{https://doi.org/10.7326/0003-4819-130-6-199903160-00016}.

\bibitem[Lee and Lee(2020)]{lee2020clinical}
Cecilia~S Lee and Aaron~Y Lee.
\newblock Clinical applications of continual learning machine learning.
\newblock \emph{The Lancet Digital Health}, 2\penalty0 (6):\penalty0
  e279--e281, 2020.

\bibitem[Liu et~al.(2020)Liu, Rivera, Moher, Calvert, and
  Denniston]{liu2020reporting}
Xiaoxuan Liu, Samantha~Cruz Rivera, David Moher, Melanie~J Calvert, and
  Alastair~K Denniston.
\newblock Reporting guidelines for clinical trial reports for interventions
  involving artificial intelligence: the consort-ai extension.
\newblock \emph{bmj}, 370, 2020.

\bibitem[Marsh et~al.(2020)Marsh, Janes, and Pepe]{marsh2020statistical}
Tracey~L Marsh, Holly Janes, and Margaret~S Pepe.
\newblock Statistical inference for net benefit measures in biomarker
  validation studies.
\newblock \emph{Biometrics}, 76\penalty0 (3):\penalty0 843--852, 2020.

\bibitem[Miller et~al.(1993)Miller, Langefeld, Tierney, Hui, and
  McDonald]{miller1993validation}
Michael~E Miller, Carl~D Langefeld, William~M Tierney, Siu~L Hui, and Clement~J
  McDonald.
\newblock Validation of probabilistic predictions.
\newblock \emph{Medical Decision Making}, 13\penalty0 (1):\penalty0 49--57,
  1993.

\bibitem[Moons et~al.(2015)Moons, Altman, Reitsma, Ioannidis, Macaskill,
  Steyerberg, Vickers, Ransohoff, and
  Collins]{moons15_trans_repor_multiv_predic_model}
Karel~G.M. Moons, Douglas~G. Altman, Johannes~B. Reitsma, John~P.A. Ioannidis,
  Petra Macaskill, Ewout~W. Steyerberg, Andrew~J. Vickers, David~F. Ransohoff,
  and Gary~S. Collins.
\newblock Transparent reporting of a multivariable prediction model for
  individual prognosis or diagnosis (tripod): Explanation and elaboration.
\newblock \emph{Annals of Internal Medicine}, 162\penalty0 (1):\penalty0
  W1--W73, 2015.
\newblock \doi{10.7326/m14-0698}.
\newblock URL \url{https://doi.org/10.7326/m14-0698}.

\bibitem[Ost(2022)]{ost2022interpretation}
David~E Ost.
\newblock Interpretation and application of the likelihood ratio to clinical
  practice in thoracic oncology.
\newblock \emph{Journal of Bronchology \& Interventional Pulmonology},
  29\penalty0 (1):\penalty0 62--70, 2022.

\bibitem[Pepe et~al.(2008{\natexlab{a}})Pepe, Feng, Huang, Longton, Prentice,
  Thompson, and Zheng]{pepe2008integrating}
Margaret~S Pepe, Ziding Feng, Ying Huang, Gary Longton, Ross Prentice, Ian~M
  Thompson, and Yingye Zheng.
\newblock Integrating the predictiveness of a marker with its performance as a
  classifier.
\newblock \emph{American journal of epidemiology}, 167\penalty0 (3):\penalty0
  362--368, 2008{\natexlab{a}}.

\bibitem[Pepe et~al.(2008{\natexlab{b}})Pepe, Feng, Janes, Bossuyt, and
  Potter]{pepe2008pivotal}
Margaret~S Pepe, Ziding Feng, Holly Janes, Patrick~M Bossuyt, and John~D
  Potter.
\newblock Pivotal evaluation of the accuracy of a biomarker used for
  classification or prediction: standards for study design.
\newblock \emph{Journal of the National Cancer Institute}, 100\penalty0
  (20):\penalty0 1432--1438, 2008{\natexlab{b}}.

\bibitem[Pepe et~al.(2016)Pepe, Janes, Li, Bossuyt, Feng, and
  Hilden]{pepe2016early}
Margaret~S Pepe, Holly Janes, Christopher~I Li, Patrick~M Bossuyt, Ziding Feng,
  and J{\o}rgen Hilden.
\newblock Early-phase studies of biomarkers: what target sensitivity and
  specificity values might confer clinical utility?
\newblock \emph{Clinical chemistry}, 62\penalty0 (5):\penalty0 737--742, 2016.

\bibitem[Rivera et~al.(2020)Rivera, Liu, Chan, Denniston, and
  Calvert]{rivera2020guidelines}
Samantha~Cruz Rivera, Xiaoxuan Liu, An-Wen Chan, Alastair~K Denniston, and
  Melanie~J Calvert.
\newblock Guidelines for clinical trial protocols for interventions involving
  artificial intelligence: the spirit-ai extension.
\newblock \emph{bmj}, 370, 2020.

\bibitem[Rosenkranz(2020)]{rosenkranz2020exploratory}
Gerd Rosenkranz.
\newblock \emph{Exploratory subgroup analyses in clinical research}.
\newblock John Wiley \& Sons, 2020.

\bibitem[Royston and Altman(2013)]{royston2013external}
Patrick Royston and Douglas~G Altman.
\newblock External validation of a cox prognostic model: principles and
  methods.
\newblock \emph{BMC medical research methodology}, 13\penalty0 (1):\penalty0
  1--15, 2013.

\bibitem[Rudin(2019)]{rudin2019stop}
Cynthia Rudin.
\newblock Stop explaining black box machine learning models for high stakes
  decisions and use interpretable models instead.
\newblock \emph{Nature Machine Intelligence}, 1\penalty0 (5):\penalty0
  206--215, 2019.

\bibitem[Simon et~al.(2003)Simon, Radmacher, Dobbin, and
  McShane]{simon2003pitfalls}
Richard Simon, Michael~D Radmacher, Kevin Dobbin, and Lisa~M McShane.
\newblock Pitfalls in the use of dna microarray data for diagnostic and
  prognostic classification.
\newblock \emph{Journal of the National Cancer Institute}, 95\penalty0
  (1):\penalty0 14--18, 2003.

\bibitem[Sox et~al.(2013)Sox, Higgins, and Owens]{sox2013medical}
H.C. Sox, M.C. Higgins, and D.K. Owens.
\newblock \emph{Medical Decision Making}.
\newblock Wiley, 2013.
\newblock ISBN 9780470658666.
\newblock URL \url{https://books.google.com/books?id=WRBszgAACAAJ}.

\bibitem[Steyerberg and Vergouwe(2014)]{steyerberg2014towards}
Ewout~W Steyerberg and Yvonne Vergouwe.
\newblock Towards better clinical prediction models: seven steps for
  development and an abcd for validation.
\newblock \emph{European heart journal}, 35\penalty0 (29):\penalty0 1925--1931,
  2014.

\bibitem[Van~Calster et~al.(2016)Van~Calster, Nieboer, Vergouwe, De~Cock,
  Pencina, and Steyerberg]{van2016calibration}
Ben Van~Calster, Daan Nieboer, Yvonne Vergouwe, Bavo De~Cock, Michael~J
  Pencina, and Ewout~W Steyerberg.
\newblock A calibration hierarchy for risk models was defined: from utopia to
  empirical data.
\newblock \emph{Journal of clinical epidemiology}, 74:\penalty0 167--176, 2016.

\bibitem[Vickers and Elkin(2006)]{vickers2006decision}
Andrew~J Vickers and Elena~B Elkin.
\newblock Decision curve analysis: a novel method for evaluating prediction
  models.
\newblock \emph{Medical Decision Making}, 26\penalty0 (6):\penalty0 565--574,
  2006.

\bibitem[Vickers et~al.(2019)Vickers, van Calster, and
  Steyerberg]{vickers2019simple}
Andrew~J Vickers, Ben van Calster, and Ewout~W Steyerberg.
\newblock A simple, step-by-step guide to interpreting decision curve analysis.
\newblock \emph{Diagnostic and prognostic research}, 3\penalty0 (1):\penalty0
  1--8, 2019.

\bibitem[Zhang et~al.(2021)Zhang, Bengio, Hardt, Recht, and
  Vinyals]{zhang2021understanding}
Chiyuan Zhang, Samy Bengio, Moritz Hardt, Benjamin Recht, and Oriol Vinyals.
\newblock Understanding deep learning (still) requires rethinking
  generalization.
\newblock \emph{Communications of the ACM}, 64\penalty0 (3):\penalty0 107--115,
  2021.

\end{thebibliography}

\end{document}